\IfFileExists{IEEEtran.cls}
  {\documentclass[conference]{IEEEtran}}
  {\documentclass[10pt,twocolumn]{article}\usepackage[margin=0.75in]{geometry}}
\usepackage[utf8]{inputenc}
\usepackage[T1]{fontenc}
\usepackage{lmodern}
\usepackage{amsmath,amssymb}
\usepackage{graphicx}
\usepackage{booktabs}
\usepackage{xcolor}
\usepackage{tikz}
\usetikzlibrary{positioning,arrows.meta,fit,backgrounds}
\usepackage{url}
\usepackage[hidelinks]{hyperref}
\newcommand{\code}[1]{\texttt{#1}}

\title{Rebuilding a Field's Open-Problem Catalog from the \\ Literature: A Two-Layer Knowledge Graph, \\ and Why Status Must Stay Proposed}
\author{%
  \IEEEauthorblockN{Madhulatha Mandarapu}
  \IEEEauthorblockA{Samyama.ai\\madhulatha@samyama.ai}
  \and
  \IEEEauthorblockN{Sandeep Kunkunuru}
  \IEEEauthorblockA{Samyama.ai\\sandeep@samyama.ai}
}

\begin{document}
\maketitle

\begin{abstract}
A research field records its open problems in surveys, keynotes and community reports. Those are
prose, written every few years, and nobody can ask them who works on a problem now or whether it
was settled last year. We argue the problems should be data, and we build the system that makes
them so: a hand-curated \emph{problem layer} carried in unchanged, under an \emph{evidence layer}
rebuilt from a bibliographic index, joined by typed, provenance-carrying links, with both layers in
one graph database and reachable from one query language. A language model does four jobs inside
this architecture --- per-paper extraction, grounding to the catalog, a per-paper status verdict,
and idea generation --- and each is constrained to a set the system offers it, so an answer that
names something outside that set is refused rather than stored.
We report a study on two curated catalogs, one of database research (1{,}053 problems) and one of
AI research (1{,}000 problems), each of which already had a graph built from the catalog alone. The
rebuilt graph answers every question the catalog-only graph answered --- 47 of 47 answerable
questions identical on databases, 34 of 35 on AI --- and answers questions the catalog-only graph
refused: who worked on a problem recently, which open problems nobody works on, what methods have
been tried, which papers from outside the catalog bear on a problem (336 problems, 770 works on
databases). The one capability we most wanted does not survive its own audit. Of 41 \emph{solves}
and \emph{refutes} verdicts hand-checked against the papers, 9 are correct, 30 are partly correct
and 2 are wrong; no \emph{refutes} verdict (0 of 34) disproves an open problem, and only 1 of 7
\emph{solves} closes the question the problem statement calls central. We report that failure in
full, and draw the design conclusion from it: a status change must be proposed with its evidence
and never applied, which makes the human reviewer part of the architecture rather than a courtesy.
We close with the data-engineering problems the vision raises: entity resolution across a curated
and a harvested layer, verifying a model's status claims, freshness under source quotas, and
evaluation without ground truth.
\end{abstract}


\section{Introduction}

Ask a database researcher which problems in the field are open and they will point you at a
community report~\cite{seattle2020,cambridge2025}. These reports are careful, broad and
influential. They are also prose, assembled every five to ten years by a committee, and they are
out of date the week they appear. Nothing in them can be asked a question. If you want to know who
works on a named problem \emph{now}, whether the problem was settled since the report was written,
or which of its open problems nobody has touched, you read the report, then you read five years of
proceedings, and then you have an opinion rather than an answer.

The same is true in machine learning, with the added difficulty that the literature doubles faster
than anyone can survey it. Mathematics has gone further: Erd\H{o}s Problems~\cite{erdosproblems}
and similar trackers keep per-problem statuses that the community updates, and a recent sweep of a
language model over 700 of its open entries resolved 13, eight of them simply by finding a paper
that had already settled the question~\cite{erdos2026}. That result is the motivation for this
work in one line: \emph{status changes hide in the existing literature}, and nobody is looking.

\paragraph{The claim.} A field's open problems should be data --- first-class entities with
identity, status and provenance --- sitting under an evidence layer that is rebuilt from the
literature, so that the questions above become queries. We are not the first to want a scholarly
knowledge graph; ORKG~\cite{orkg}, Semantic Scholar~\cite{s2} and OpenAlex~\cite{openalex} model
papers, authors and contributions at a scale we cannot approach. What none of them models is a
\emph{problem with a status that evidence can change}. And what the mathematics trackers lack is a
bibliographic graph underneath, typed links into it, and a query language over the pair.

\paragraph{What this paper adds.} Our contributions are:

\begin{enumerate}
\item \textbf{Five requirements} for a living open-problem graph (Section~\ref{sec:vision}), each
      stated with the failure it prevents, derived from defects we hit in building one.
\item \textbf{A two-layer architecture} (Section~\ref{sec:arch}): a curated problem layer carried
      in unchanged under a rebuilt evidence layer, joined by \code{SAME\_AS} and
      \code{EVIDENCE\_FOR} edges, with the strict rule that the two layers never share a node
      label.
\item \textbf{A pipeline} (Section~\ref{sec:pipeline}) in which every language-model stage is
      constrained to a set the system offers it, and a refusal is a first-class outcome. 0 of 104
      generated research ideas across two runs named a problem they were not offered.
\item \textbf{A study on two fields} (Sections~\ref{sec:study}--\ref{sec:results}): the rebuilt
      graph is non-regressive on the catalog-only graph's own benchmark, and it answers questions
      that graph refused.
\item \textbf{A negative result} (Section~\ref{sec:verdicts}) on the component the vision most
      depends on. Hand-checking all 41 \emph{solves} and \emph{refutes} verdicts the system
      produced gives 9 correct, 30 partly correct, 2 wrong. The failure is systematic, not noisy:
      the dominant pattern (9 of 41) is a paper showing that current methods fail, read as
      refuting an open ``can it be done'' question.
\item \textbf{A research agenda} (Section~\ref{sec:agenda}) of the data-engineering problems this
      raises, each with an observation from the study rather than a guess.
\end{enumerate}

\paragraph{What this paper does not claim.} We do not claim the system can tell whether a problem
is solved. Section~\ref{sec:verdicts} is the evidence that it cannot, yet. We do not claim
coverage: the evidence layer reaches 579 of 1{,}053 database problems and 592 of 1{,}000 AI
problems. We do not claim the catalogs are authoritative; they are two catalogs we built, and the
architecture is indifferent to who curates them. And every fact in the graph comes from titles,
abstracts and metadata, not full text.

\section{Requirements}
\label{sec:vision}

The five requirements below are not a wish list. Each one is a defect we hit, generalised.

\paragraph{R1 --- Stable identity, separately for each layer.} A problem, a paper and a person each
need an identifier that survives a rebuild. The failure it prevents is subtle: our first rebuild
gave the harvested bibliographic records the same node labels (\code{Paper}, \code{Person},
\code{Institution}, \code{Venue}) as the curated catalog. Nothing errored. A query that counted
``papers the catalog cites'' returned 8{,}769 instead of 2{,}679, and a query for name variants
returned a \code{Person} node whose name was a five-author list. The fix was to rename every label
in the evidence layer (\code{Work}, \code{Author}, \code{OpenAlexInstitution},
\code{OpenAlexSource}) and to make a shared label a build-time error. See
Section~\ref{sec:collision}.

\paragraph{R2 --- Provenance on every edge and node.} Every node carries a \code{source} field:
\code{openalex} for a bibliographic fact, \code{extraction} for something a model read out of an
abstract, \code{web} for something it found by searching, \code{synthesis} for something it wrote.
Without this, a reader cannot tell a citation from a guess, and we could not have written
Section~\ref{sec:verdicts} at all.

\paragraph{R3 --- Freshness as a rebuild, not an update.} The evidence layer is derived, so it is
rebuilt rather than patched; the catalog is curated, so it is versioned. The two are joined at
build time. This is what makes ``which evidence appeared since the catalog was last edited'' a
query: on AI, 71 works from 2026, 66 of which the catalog does not contain.

\paragraph{R4 --- A status change is proposed, never applied.} The system writes a suggestion with
its evidence and its reasoning onto the graph. It does not change the catalog. Before the audit in
Section~\ref{sec:verdicts} this was a conservative default; after it, it is the single most
important thing in the design.

\paragraph{R5 --- One query language over both layers.} Both layers live in one graph database and
are reached with openCypher. The benchmark in Section~\ref{sec:study} is possible only because the
two graphs under comparison answer the same query text.

\section{Architecture}
\label{sec:arch}

\begin{figure}[t]
\centering
\resizebox{\columnwidth}{!}{%
\begin{tikzpicture}[
  font=\footnotesize,
  box/.style={draw, rounded corners=2pt, align=center, inner sep=3pt, minimum height=7mm, minimum width=24mm},
  cat/.style={box, fill=blue!6, draw=blue!50},
  ev/.style={box, fill=orange!8, draw=orange!60},
  lay/.style={draw, dashed, rounded corners=3pt, inner sep=7pt},
  link/.style={-{Latex[length=1.6mm]}, thick},
  bridge/.style={link, densely dashed, draw=black!70},
  lbl/.style={font=\scriptsize, inner sep=1.5pt, fill=white},
]
\node[cat] (prob) at (0,0)    {\code{Problem}\\[-1pt]\scriptsize status, rationale};
\node[cat] (cpap) at (3.9,0)  {\code{Paper}\\[-1pt]\scriptsize catalog-cited};
\node[cat] (cper) at (7.8,0)  {\code{Person}\\[-1pt]\scriptsize named in catalog};
\draw[link, blue!60] (prob) -- node[lbl] {\code{CITES}} (cpap);
\draw[link, blue!60] (cpap) -- node[lbl] {catalog edge} (cper);
\begin{scope}[on background layer]
  \node[lay, draw=blue!50, fit=(prob)(cpap)(cper)] (catbox) {};
\end{scope}
\node[lbl, fill=none, anchor=south west, blue!60!black] at (catbox.north west)
  {\itshape problem layer --- curated, versioned, carried in unchanged};

\node[ev] (pp)   at (0,-3.6)   {\code{PaperProblem}\\[-1pt]\scriptsize extracted};
\node[ev] (work) at (3.9,-3.6) {\code{Work}\\[-1pt]\scriptsize OpenAlex record};
\node[ev] (auth) at (7.8,-3.6) {\code{Author}\\[-1pt]\scriptsize OpenAlex author};
\draw[link, orange!70] (work) -- node[lbl] {\code{RAISES}} (pp);
\draw[link, orange!70] (work) -- node[lbl] {\code{AUTHORED}} (auth);
\begin{scope}[on background layer]
  \node[lay, draw=orange!60, fit=(pp)(work)(auth)] (evbox) {};
\end{scope}
\node[lbl, fill=none, anchor=north west, orange!50!black] at (evbox.south west)
  {\itshape evidence layer --- rebuilt from OpenAlex and model extraction};

\draw[bridge] (pp.north)   -- node[lbl, align=center]
  {\code{EVIDENCE\_FOR}\\verdict, quote, confidence} (prob.south);
\draw[bridge] (work.north) -- node[lbl] {\code{SAME\_AS}} (cpap.south);
\draw[bridge] (auth.north) -- node[lbl] {\code{SAME\_AS}} (cper.south);
\end{tikzpicture}}
\caption{The two layers and the bridge between them. No node label appears in both layers; a shared
label is a build-time error (Section~\ref{sec:collision}). Every bridge edge carries how it was
made, and \code{EVIDENCE\_FOR} additionally carries the verdict, the quote it rests on and the
model's confidence.}
\label{fig:arch}
\end{figure}
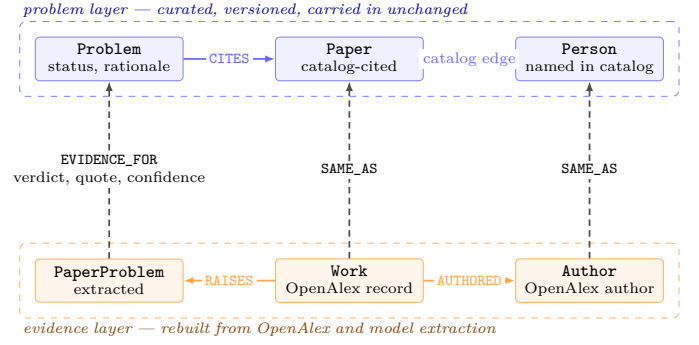

Figure~\ref{fig:arch} shows the shape. Three decisions in it are load-bearing.

\paragraph{The catalog is carried in, not merged.} The curated layer enters the graph byte for byte:
same ids, same statuses, same text. Nothing in the pipeline writes to it. This is what lets us run
the catalog-only graph's own benchmark against the rebuilt graph and read any difference as a
defect (Section~\ref{sec:nonregression}).

\paragraph{No label is shared between layers.} A graph database will happily let two populations
share a label, and then every unlabelled or singly-labelled pattern silently spans both. The build
now refuses to start if an evidence-layer label collides with a catalog label.

\paragraph{The bridge is explicit and typed.} Catalog papers are matched to bibliographic records by
DOI, then by arXiv identifier, then by title and year; catalog people are matched through the
papers they authored, and only when exactly one candidate matches. The result is a \code{SAME\_AS}
edge carrying which rule fired, so a reader can discount a title-match differently from a DOI
match. \code{EVIDENCE\_FOR} runs from a work to a problem and carries the verdict, the quoted
sentence it rests on and a confidence.

A consequence worth naming: catalog name variants are \emph{linked, not merged}. Three catalog
nodes (``Neumann'', ``T. Neumann'', ``Thomas Neumann'') resolve to one \code{Author}; the three
catalog nodes stay. Queries that want the person traverse \code{SAME\_AS}; queries that want the
catalog's own text do not. Merging would have been destructive and irreversible, and we have been
burned by exactly that in a different graph.

\section{The Pipeline}
\label{sec:pipeline}

The build runs nine stages. Five are deterministic (bibliographic harvest, bridge, graph build,
snapshot export, benchmark); four use a language model. The interesting design content is in how
the four are constrained.

\begin{table}[t]
\centering\footnotesize
\caption{Pipeline stages. ``Offered set'' is what the model may name; anything else is refused.}
\label{tab:stages}
\begin{tabular}{@{}llp{2.5cm}@{}}
\toprule
Stage & Kind & Offered set \\
\midrule
1 resolve author ids   & model + web  & candidates from the index \\
2 harvest works        & index        & --- \\
3 expand (refs, citers)& index        & --- \\
4 extract per paper    & model        & free text, then canonicalised \\
5 profile people       & model + web  & --- \\
6 map the domain       & model + web  & --- \\
7 canonicalise         & model        & the names already extracted \\
8 ground + status      & model        & the catalog's own problems \\
9 generate ideas       & model        & the open problems offered \\
\bottomrule
\end{tabular}
\end{table}

\paragraph{Extraction (stage 4).} Title and abstract of eight papers per call; out come problems,
methods, materials, properties, results, tools, applications and stated future work, each becoming
a node with \code{source=extraction}. The model is reading, not judging.

\paragraph{Canonicalisation (stage 7).} All extracted names of one kind go in; out comes a synonym
map and \code{BROADER} parents. The offered set is the set of names already extracted, so this
stage cannot invent a concept.

\paragraph{Grounding and the status check (stage 8).} This is where the two layers meet. For each
extracted per-paper problem the model is given the catalog's problems and must answer with one of
them, at the relation \emph{same} or \emph{narrower}, or decline. Only then, for a catalog problem
that has evidence, does the status check run: for each of up to eight evidence papers it returns one
of \emph{advances}, \emph{restates}, \emph{solves}, \emph{refutes} or \emph{unrelated}, with a
quote and a confidence. A \emph{solves} or \emph{refutes} verdict may yield a \emph{suggested}
status, written onto the problem as a suggestion with its rationale. The catalog is not touched.

\paragraph{Idea generation (stage 9).} The model is offered the open problems of the catalog and the
domain map, and may propose ideas only against problems in that offered set. An idea naming anything
else is rejected at parse time. Across the two runs reported here, 104 ideas were generated and 0
were rejected on this rule --- which says the constraint is satisfiable, not that it is unnecessary.

\paragraph{Implementation.} About 3{,}200 lines of Python. The graph is loaded over HTTP into a
graph--vector database (\code{samyama-graph}, Apache-2.0, run from a public container image) with
\code{UNWIND \$rows ... MERGE ... SET n += r} and an index on \code{id} per label; the run then
exports a portable snapshot. Every stage caches its output in the run directory --- index responses,
each model call, the extraction and canonicalisation results --- so a re-run pays only for what did
not finish, and deleting one file redoes one stage. The model stages go through a batch API at half
price where they can; on the AI run, 3{,}690 of 3{,}698 calls did.

\section{Study Design}
\label{sec:study}

We had an unusual opportunity: two curated catalogs that \emph{already} had a graph built from the
catalog alone, each with its own hand-written benchmark of scope questions. Call the catalog-only
graph V1 and the rebuilt two-layer graph V3. Both are built from the same catalog commit. The
comparison is therefore not system-versus-baseline but the same catalog with and without an
evidence layer beneath it, on queries written before this architecture existed.

\begin{table}[t]
\centering\footnotesize
\caption{The two fields. V1 is the catalog-only graph; V3 adds the evidence layer. Catalog commits
and bench files are recorded in the artifact.}
\label{tab:fields}
\begin{tabular}{@{}lrr@{}}
\toprule
 & Databases & AI \\
\midrule
Catalog problems            & 1{,}053  & 1{,}000 \\
Catalog papers              & 2{,}679  & 4{,}932 \\
Catalog people              & 4{,}454  & 15{,}102 \\
Seed researchers            & 40       & 40 \\
\addlinespace
V1 nodes                    & 20{,}211 & 29{,}707 \\
V1 edges                    & 53{,}721 & 62{,}842 \\
V3 nodes                    & 78{,}475 & 120{,}972 \\
V3 edges                    & 313{,}878& 463{,}344 \\
V3 works in evidence layer  & ---      & 7{,}877 \\
\addlinespace
Benchmark questions         & 48       & 35 \\
Run date                    & 2026-10-20 & 2026-10-23 \\
\bottomrule
\end{tabular}
\end{table}

We ask three things of the comparison.

\begin{enumerate}
\item \textbf{Non-regression.} Does V3 still answer what V1 answered, identically? Any difference
      is a defect in V3, because the catalog did not change.
\item \textbf{New answers.} The questions V1's own documentation marks as answered in part,
      unanswered or out of scope: rewritten against V3's schema, with the same parameters, do they
      now return something, and is that something right?
\item \textbf{Audit.} Where the system makes a judgement rather than a join, how good is the
      judgement?
\end{enumerate}

The third is the part that most systems papers skip. We ran it on every \emph{solves} and
\emph{refutes} verdict the two runs produced --- 41 in total, not a sample --- and report it in
Section~\ref{sec:verdicts}.

\section{Results}
\label{sec:results}

\subsection{Non-regression}
\label{sec:nonregression}

On databases, 47 of the 48 benchmark questions have a Cypher query (the 48th needs vector search).
\textbf{All 47 return identical answers on V1 and V3}, compared order-insensitively on rows,
columns and the first twenty records.

On AI, \textbf{34 of 35 queries return the same rows}. 26 are identical row for row; 8 differ only
in the order of rows that tie on the sort key; one (``papers cited by most problems'') differs in a
single row because three papers tie at 35 citing problems for the last two places under
\code{LIMIT 10}, and the two graphs keep a different pair. Without the limit both graphs return the
same 11 papers. Two queries that must return nothing return nothing on both.

This is a weak result by construction and we state it as such: the original queries read only the
catalog layer, so they cannot show what the evidence layer adds. Their job is to prove it adds
nothing \emph{wrong}. It took a fix to get here (Section~\ref{sec:collision}).

\subsection{What the evidence layer adds}

\begin{table*}[t]
\centering\footnotesize
\caption{Questions the catalog-only graph answered in part, refused, or listed as out of scope,
re-asked against the rebuilt graph with the same parameters. Full per-question tables, including
the ones not attempted and why, are in the artifact.}
\label{tab:newanswers}
\scriptsize
\begin{tabular}{@{}lp{4.6cm}p{3.6cm}p{6.6cm}@{}}
\toprule
Field & Question & V1 & V3 \\
\midrule
DB & Papers after 2015 bearing on the problems    & 787 catalog papers      & 873 evidence works, 480 new to the catalog \\
DB & Which catalog papers a bibliographic index verifies & DOI on 1{,}800 of 2{,}679 & 2{,}278 matched; 386 have neither a DOI nor a match \\
DB & References no bibliographic source verifies  & 15                      & the same 15 (corroborated independently) \\
DB & Researchers with an ORCID                    & 0 rows                  & 2{,}181 via their matched author \\
DB & Name variants of one researcher              & 3 separate nodes        & all three resolve to one author \\
DB & Open problems nobody works on                & 72                      & 38, counting evidence works as work \\
DB & Open problems citing nothing after 2015      & 86                      & 70 \\
DB & Total citations of a paper                    & refused                 & citation counts on all 2{,}278 matched papers \\
DB & Has the problem been solved since?            & refused                 & 4 \emph{solves} verdicts, 2 suggested changes --- see Section~\ref{sec:verdicts} \\
DB & Who funds work on the problem                 & refused                 & 3 funders; funder names on 567 catalog papers \\
\addlinespace
AI & Methods tried on a problem                   & no method data          & 23 methods from 10 of its 12 cited papers \\
AI & Methods used for the most problems           & unanswered              & literature review 125, scaling analysis 111, SFT 88, LoRA 43 \\
AI & Problems whose papers use a given method     & unanswered              & 61 problems for LoRA \\
AI & A researcher's top co-authors                & 10 shared catalog papers& 32, 26, 17 over the run's works \\
AI & Researchers with an ORCID                    & 0                       & 7{,}320 \\
AI & Recent work on the problems                  & newest 2025, none 2026  & 71 works from 2026, 66 not catalog papers \\
AI & Open problems with no lab                    & 54                      & 54, of which 23 have no evidence work either \\
AI & Citation counts                              & out of scope            & on all 4{,}630 matched papers \\
AI & Has any problem been solved or refuted?      & refused                 & 3 \emph{solves}, 32 \emph{refutes}, 2 suggested changes \\
\bottomrule
\end{tabular}
\end{table*}

Table~\ref{tab:newanswers} is the substance of the comparison. Reading it with the per-question
notes in the artifact, we grade the outcomes by hand:

\begin{itemize}
\item \textbf{Databases.} V1 marked 20 questions as answered in part and refused 3. We attempted 11
      of those 23; the other 12 ask about content \emph{inside} the catalog (bounds, concept
      definitions, which system currently leads), which an evidence layer does not supply. Of
      the 11: 8 got a better answer, 1 got the same answer now corroborated by an independent
      source, 1 did not change, and 1 (``has it been solved'') is answered but unverified.
\item \textbf{AI.} The AI catalog grades no answer as partial, so we assembled 18 questions from
      its own list of unanswered questions and known limits. 16 were attempted: 5 better, 4 went
      from refused to answered, 3 from refused to partly answered, 1 mixed, 1 unchanged by design,
      2 genuinely unchanged.
\end{itemize}

The honest summary of the ``mixed'' row is worth more than the eight wins. Asked which researchers
are linked to the most problems, V1 returns four names; V3 merges their name variants and so ranks
them better --- and drops one of them entirely, because that researcher's catalog node has no
matched author. The bridge improves the ranking and loses a seed researcher in the same query.
Section~\ref{sec:bridge} says why.

Beyond the question list, the evidence layer adds a population that did not exist before:

\begin{table}[t]
\centering\footnotesize
\caption{Evidence reaching the catalog's problems.}
\label{tab:evidence}
\begin{tabular}{@{}lrr@{}}
\toprule
 & Databases & AI \\
\midrule
Catalog problems                        & 1{,}053 & 1{,}000 \\
\dots with at least one evidence work   & 579     & 592 \\
Evidence works                          & 1{,}773 & 1{,}969 \\
\code{EVIDENCE\_FOR} links              & 1{,}843 & 2{,}013 \\
\dots with works from outside the catalog & 336 & 187 \\
\dots those works                       & 770     & 300 \\
\dots with works they do not cite       & ---     & 360 \\
\addlinespace
Per-paper problems extracted        & 5{,}058 & 6{,}885 \\
\dots grounded to a catalog problem     & 1{,}265 & 1{,}311 \\
\quad as the same problem               & 138     & 145 \\
\quad as a narrower problem             & 1{,}127 & 1{,}166 \\
Grounded share                          & 25.0\%  & 19.0\% \\
\bottomrule
\end{tabular}
\end{table}

Two readings of Table~\ref{tab:evidence} matter. First, ``problems with works from outside the
catalog'' is the capability the whole architecture exists for: 336 database problems have 770 papers
bearing on them that the catalog's curators never cited. Second, the grounded share is low ---
19.0\% on AI, 25.0\% on databases. Four fifths of what the model reads out of an abstract does not
correspond to anything in the curated catalog. That is partly the catalog's narrowness and partly
the extraction's granularity, and we cannot currently separate the two.

\subsection{The bridge, and what it costs}
\label{sec:bridge}

\begin{table}[t]
\centering\footnotesize
\caption{Bridging the curated layer to the harvested one. On AI, exactly 1 catalog paper carries a
DOI, so the DOI rule --- the most reliable one --- fires 0 times.}
\label{tab:bridge}
\begin{tabular}{@{}lrr@{}}
\toprule
 & Databases & AI \\
\midrule
Catalog papers                 & 2{,}679 & 4{,}932 \\
\dots matched                  & 2{,}278 & 4{,}630 \\
\quad by DOI                   & ---     & 0 \\
\quad by arXiv id              & ---     & 3{,}026 \\
\quad by title and year        & ---     & 1{,}604 \\
\dots unmatched                & 401     & 302 \\
Works claimed by two catalog papers & --- & 21 \\
\addlinespace
Catalog people                 & 4{,}454 & 15{,}102 \\
\dots linked                   & 3{,}087 & 11{,}657 \\
\dots unlinked                 & 1{,}367 & 3{,}445 \\
Distinct authors matched       & 2{,}488 & 7{,}807 \\
Authors covering several catalog nodes & --- & 2{,}732 \\
\quad catalog nodes on those   & ---     & 6{,}582 \\
\bottomrule
\end{tabular}
\end{table}

Table~\ref{tab:bridge} is the most useful table in the paper for anyone building this. Three things
in it were surprises.

\textbf{The best matching rule can fire zero times.} On the AI catalog, one paper out of 4{,}932
carries a DOI, because the field publishes to arXiv and the curators recorded arXiv ids. So 3{,}026
matches came from arXiv ids and 1{,}604 from title and year --- the weakest rule carrying a third of
the bridge. A system that implements only DOI matching would have bridged one paper.

\textbf{Precision on people costs recall, visibly.} We link a catalog person to an author only when
exactly one candidate matches. One seed researcher in the AI run --- the person ranked third by
problem count in V1 --- is split across several author ids upstream, with seven of his catalog
papers under one id and two under another, so no single candidate matches and he is left unlinked.
That is the rule working as designed, and it drops a seed researcher out of a headline query.
3{,}445 of 15{,}102 catalog people are unlinked.

\textbf{Linking is not merging, and the distinction is measurable.} 2{,}732 authors in the AI run
each link more than one catalog node, covering 6{,}582 catalog nodes between them. Those 6{,}582
nodes stay. A merge would have collapsed them and silently rewritten every count the catalog's own
benchmark asserts.

\subsection{Status verdicts: the component that failed its audit}
\label{sec:verdicts}

The status check is the reason to build any of this. It ran on 572 database problems over 1{,}653
papers and on 578 AI problems, eight papers per problem.

\begin{table}[t]
\centering\footnotesize
\caption{Verdicts produced by the status check. \emph{none} is a link the check never reached,
because it reads at most eight papers per problem.}
\label{tab:verdicts}
\begin{tabular}{@{}lrr@{}}
\toprule
Verdict & Databases & AI \\
\midrule
advances   & 893 & 1{,}062 \\
restates   & 403 & 235 \\
unrelated  & 351 & 416 \\
\emph{solves}  & 4 & 3 \\
\emph{refutes} & 2 & 32 \\
none (not reached) & 190 & 265 \\
\addlinespace
Status changes suggested & 2 & 2 \\
\bottomrule
\end{tabular}
\end{table}

Table~\ref{tab:verdicts} looks encouraging: the system is conservative, calling \emph{solves} 7
times in 3{,}856 graded links and suggesting only 4 status changes. We then checked all 41
\emph{solves} and \emph{refutes} verdicts by hand against the problem statement, the status
rationale, the work's title and abstract, and the quote the verdict rests on. Full texts were not
read; the graph was not changed.

\begin{table}[t]
\centering\footnotesize
\caption{Hand check of every \emph{solves} and \emph{refutes} verdict both runs produced ($n=41$,
not a sample). \emph{correct}: the paper does what the verdict says to a claim or question in the
problem statement. \emph{partly}: it bears on the problem but advances it, settles a sub-case, or
contradicts an adjacent or rival claim. \emph{wrong}: it does not do what the verdict says.}
\label{tab:handcheck}
\begin{tabular}{@{}llrrrr@{}}
\toprule
Field & Verdict & $n$ & correct & partly & wrong \\
\midrule
AI & \emph{solves}  & 3  & 0 & 3  & 0 \\
AI & \emph{refutes} & 32 & 8 & 23 & 1 \\
DB & \emph{solves}  & 4  & 1 & 3  & 0 \\
DB & \emph{refutes} & 2  & 0 & 1  & 1 \\
\midrule
\multicolumn{2}{@{}l}{All} & 41 & 9 & 30 & 2 \\
\bottomrule
\end{tabular}
\end{table}

Table~\ref{tab:handcheck} is the paper's central negative result. Stated plainly:

\begin{itemize}
\item \textbf{No \emph{refutes} verdict (0 of 34) disproves an open problem.} The pipeline's
      \emph{refutes} means ``contradicts a claim on the problem page''. Read that way, 8 of 32 AI
      \emph{refutes} are correct; read as ``this problem is closed negatively'', none are.
\item \textbf{Only 1 of 7 \emph{solves} closes the question the statement calls central.} The other
      six settle a sub-case: a tight privacy bound that holds only for smooth convex losses on a
      bounded domain while the problem's own title names the non-convex case; a contamination fix
      demonstrated at BERT scale while the problem asks about large-model pretraining.
\item \textbf{Nothing was silently wrong in the graph}, because R4 holds. Four suggestions were
      written; no status was changed.
\end{itemize}

The failures are systematic, which is the useful part. Classifying all 41 by failure pattern:

\begin{table}[t]
\centering\footnotesize
\caption{Failure patterns across the 41 hand-checked verdicts.}
\label{tab:patterns}
\begin{tabular}{@{}lr@{}}
\toprule
Pattern & $n$ \\
\midrule
``methods fail'' read as refuting an open ``can it be done?'' & 9 \\
settles a sub-case only                                                & 6 \\
evidence from an adjacent or restricted setting                        & 6 \\
answers or advances a sub-question, labelled \emph{refutes}    & 5 \\
refutes a rival explanation or prior result, not the problem   & 4 \\
source paper of the phenomenon, labelled \emph{refutes}                & 1 \\
a performance comparison read as a refutation                          & 1 \\
\bottomrule
\end{tabular}
\end{table}

The dominant pattern in Table~\ref{tab:patterns} is a logic error, not a reading error: a paper
showing that today's methods fail at a task is evidence \emph{about} an open ``is it possible''
question, never a refutation of it. Nine of 41 verdicts make that one mistake.

A second pair of patterns (\emph{settles a sub-case only}, $n=6$, and \emph{refutes a rival
explanation or prior result, not the problem}, $n=4$) has a direct parallel in formal
mathematics. In the October 2026 Lean formalization of the classification of finite simple
groups~\cite{axiomcfsg2026}, candidate agent proofs repeatedly established a milestone's target
conclusion under additional unproved assumptions --- proving a conditional sub-case while leaving
the milestone open --- and an explicit counterexample ($\mathrm{PSL}_2(49)$) to a closure
assertion in Walter's 1986 Unbalanced Group Theorem route refuted that proof step without
refuting the theorem itself. Even when the evidence is a deductive proof, settling a restricted
case and refuting an intermediate claim are distinct from closing the problem.

The second finding is about the problem statements rather than the model. The AI catalog's problem
statements are one paragraph (309--1{,}060 characters) and most pose a question rather than assert a
claim --- so \emph{refutes} often has no target to attach to. Five of the eight correct AI
\emph{refutes} are on the minority of problems whose statement names the claim under examination.
\textbf{Writing the claim down is what makes it refutable.} This is a requirement on the curator,
not on the model, and we did not anticipate it.

Confidence was mildly informative and not usable as a gate. Mean confidence on the AI verdicts was
0.73 for correct, 0.6 for partly correct and 0.4 for the single wrong one; on databases, 0.8, 0.61
and 0.4. The ordering is right and $n=41$ is far too small to set a threshold on; we report it as
observed, not as a calibration result.

\subsection{Grounding to a shared vocabulary}

Separately from the catalog, the AI run grounded its extracted concepts to a merged
computer-science vocabulary of 13{,}551 classes (CSO and the ACM classification), to test whether
the extraction's output can be placed in a vocabulary nobody in this project wrote.

\begin{table}[t]
\centering\footnotesize
\caption{Share of extracted concepts placed at \emph{same} or \emph{narrower} against a
13{,}551-class computer-science vocabulary (AI run). 9 answers were refused for naming a class not
offered.}
\label{tab:ontology}
\begin{tabular}{@{}lrr@{}}
\toprule
Kind & Extracted & Placed \\
\midrule
applications      & 4{,}665  & 38.6\% \\
subjects          & 7{,}564  & 31.6\% \\
problems          & 6{,}885  & 28.5\% \\
methods           & 14{,}198 & 25.4\% \\
metrics           & 5{,}198  & 12.9\% \\
datasets, tools   & 4{,}431  & 5.9\% \\
\midrule
all               & 42{,}941 & 24.9\% \\
\bottomrule
\end{tabular}
\end{table}

Table~\ref{tab:ontology} says something we did not expect to have to say: a general
computer-science vocabulary places a quarter of what a current AI paper talks about. Datasets and
tools are the floor at 5.9\%, which is unsurprising --- vocabularies do not enumerate benchmarks
--- but methods at 25.4\% is a gap in the vocabulary, not in the extraction. The named entities of
2024--2026 machine learning are largely absent from the shared vocabularies of the field.

\subsection{Cost}

\begin{table}[t]
\centering\footnotesize
\caption{Model cost of one full build. Tokens are from the run's usage log. No dollar figure is
computed: the pipeline carries no price table, and quoting one would be a number with no artifact
behind it.}
\label{tab:cost}
\begin{tabular}{@{}lrr@{}}
\toprule
 & Databases & AI \\
\midrule
Model calls            & 2{,}808 & 3{,}698 \\
\dots through a batch API & 2{,}799 & 3{,}690 \\
Input tokens (M)       & 23.6 & 28.6 \\
Output tokens (M)      & 6.5  & 9.6 \\
Web searches           & 431  & 417 \\
\addlinespace
Snapshot size (MB)     & 16.9 & 16.8 \\
\bottomrule
\end{tabular}
\end{table}

Table~\ref{tab:cost} sizes the thing. A field of a thousand problems and a few thousand papers
costs a few thousand model calls and about 30M input tokens to rebuild, almost all of it eligible
for batch pricing, and the result is a 17\,MB portable snapshot. The deep model is used sparingly:
54 of the 3{,}698 AI calls, consuming 7.0M input tokens between them. The binding constraint in
practice was not the model but the bibliographic index's free-tier daily quota, which stopped one
run outright rather than returning partial data.

\section{Three Defects, Reported in Full}
\label{sec:collision}

\paragraph{A shared label silently doubled a count.} Described in R1. What makes it worth a
paragraph is the failure mode: no error, no warning, a plausible number. A query counting
``papers the catalog cites'' returned 8{,}769 where the catalog has 2{,}679, because the harvested
records wore the same \code{Paper} label. We found it only because an older benchmark asserted the
right number. The 8{,}769 is quoted from that build's run report; the build no longer exists, so we
could not recompute it. Without V1 to compare against, the first build would have shipped. Generalised: when
you add a harvested layer under a curated one, \textbf{a label collision is a correctness bug that
presents as a larger, more impressive graph.}

\paragraph{An upstream abstract belonged to a different paper.} The index's record for a
well-known 2020 scaling-laws paper carries an abstract about a wholly unrelated topic; the cached
raw response carries the same text, so the error is upstream of us. Extraction duly read six
unrelated methods out of it, and because 54 catalog problems cite that paper, those six methods
ranked fifth to tenth in a first draft of ``methods used for the most problems''. The mitigation is
a frequency floor --- a method must appear in at least three papers to be ranked --- which is a
mitigation, not a fix. A heuristic scan found 266 of 7{,}040 works whose abstract shares no content
word with its title (author lists, placeholder text, genuinely wrong abstracts); that is a derived
upper bound on the problem, not a verified count. Note what it means for the audit: a bad abstract
propagates to a verdict, and in this run the known-bad paper carried no \emph{solves} or
\emph{refutes} verdict, which is luck, not design.

\paragraph{A precision rule dropped a seed researcher.} Described in Section~\ref{sec:bridge}. We
keep the rule and report the loss, because the alternative --- linking on a best guess --- puts a
wrong co-authorship into a graph whose whole value proposition is provenance.

\section{Limitations}

\begin{itemize}
\item \textbf{The audit's judge is not independent.} The 41 verdicts were judged with model
      assistance by one of the authors, from titles, abstracts and quotes, without reading full
      texts. It is a strict enough test to produce a negative result; it is not a
      domain-expert adjudication, and the \emph{correct}/\emph{partly} boundary is a judgement
      call we made.
\item \textbf{No ground truth for status.} Nobody has a labelled set of ``problems solved between
      2015 and 2026''. Until someone builds one, every evaluation of a status check is an audit of
      its outputs, which can measure precision and cannot measure recall. We do not know how many
      genuine solutions the system missed.
\item \textbf{Two fields, one curator.} Both catalogs were built by us, which risks a schema that
      fits our own curation habits.
\item \textbf{Abstracts, not full text.} Every extracted fact rests on a title, an abstract and
      metadata.
\item \textbf{The improvement grades are ours.} The AI catalog grades no answer as partial, so
      the 18-question set in Table~\ref{tab:newanswers} is our own reading of its scope notes and
      stated limits, and the better/same/no-change grades in both fields are ours, read from the
      returned rows. No automatic checker covers them.
\item \textbf{The run logs are not public.} The numbers in this paper come from committed summary
      files; the underlying per-call logs are large and contain third-party abstracts, and are not
      published.
\item \textbf{One benchmark per field, written by the catalog's own authors.} The non-regression
      result inherits whatever that benchmark does not ask.
\end{itemize}

\section{A Research Agenda}
\label{sec:agenda}

Each item below is a data-engineering problem, with the observation from this study that makes it
concrete.

\paragraph{(a) Entity resolution across a curated and a harvested layer.} Not record linkage in
general: linkage where one side is small, hand-made and authoritative about its own text, and the
other is large, noisy and authoritative about the world. Our observation: the DOI rule fired 0
times on one of two catalogs, the precision rule leaves 3{,}445 of 15{,}102 people unlinked, and a
label collision inflated a count by $3.3\times$ without an error.

\paragraph{(b) Verifying a status claim.} The open question this paper leaves is not ``can a model
read a paper'' but ``can a system tell the difference between settling a question and settling a
sub-case''. 6 of our 41 verdicts fail exactly there, and 9 more confuse evidence about a question
with an answer to it. We suspect this needs the problem statement to carry a formal or
semi-formal claim and an explicit sub-case decomposition --- analogous to decomposing a theorem
into 108 milestones and 380 tasks with explicit hypotheses before checking whether the pieces
close the target~\cite{axiomcfsg2026} --- which makes it a data-modelling problem rather than a
prompting problem.

\paragraph{(c) Freshness under a source quota.} The rebuild is the freshness mechanism, and the
rebuild is rate-limited by an external index whose free tier is a per-IP daily budget. Incremental
rebuild, per-problem prioritisation and a cost model for ``what is worth re-checking this month''
are all open.

\paragraph{(d) Provenance and contestability.} Every edge records how it was made, which is
necessary and not sufficient. A researcher who finds a wrong \code{SAME\_AS} on their own name, or
disagrees with a \emph{refutes} verdict on their own problem, needs a way to dispute it that
survives the next rebuild. We have no mechanism for this.

\paragraph{(e) Evaluation without ground truth.} Our non-regression test works because a prior
graph existed with its own benchmark. That is a lucky accident of our history, not a method. The
field needs an evaluation design for a system whose output is a proposal about the state of
knowledge.

\paragraph{(f) Curation and governance.} The catalog is the authority in this architecture, and in
our study its curators are its authors. The mathematics trackers took the opposite route --- a
community updating statuses by pull request --- and that is the part of their design we would most
like to copy.

\section{Related Work}

\paragraph{Open-problem trackers.} Mathematics has them. Erd\H{o}s
Problems~\cite{erdosproblems} carries roughly 1{,}200 problems with an explicit status, updated by
pull request, and a recent language-model sweep over its open entries resolved 13, eight of them by
locating existing literature~\cite{erdos2026}. ProofAtlas publishes 500 open problems with statuses
that include partially-solved and with dated status sources; Open Problem Garden is a manually
maintained wiki. At the formal end, the Lean formalization of the classification of finite simple
groups~\cite{axiomcfsg2026} mapped 715 papers and books into a blueprint of 108 milestones and
about 380 tasks above 2.05 million kernel-checked declarations, surfacing 34 deviations from the
published record. These are the motivation for this work, not a competitor to it: they establish
both the demand and the premise that status changes and literature gaps are already sitting in
published work. Where a formal proof has Lean's kernel to verify that sub-tasks close a milestone,
an empirical or mixed computer-science field does not; what the mathematics trackers also lack is a
bibliographic graph beneath the problems, typed links into it, or a query language over the pair.
Our contribution is squarely that missing layer, for a computer-science field.

\paragraph{Scholarly knowledge graphs.} ORKG~\cite{orkg} models papers, contributions and research
problems with comparisons over them, and is the closest prior art; it does not give a problem a
status that evidence can change. Semantic Scholar~\cite{s2} and OpenAlex~\cite{openalex} are the
bibliographic substrate --- we build on the latter --- and model no problem status by design.
Papers with Code tracked best-reported scores per task and dataset, which is a number rather
than a status, and shut down in 2025.

\paragraph{Mining problems from text.} Work on extracting challenges and future directions from
papers establishes that the extraction is feasible, and stops at extraction: the output is a list
or a search index, with no persistent entity, no status and no link to a curated catalog.

\paragraph{Idea-generation agents.} Systems that propose research ideas with a novelty
check~\cite{aiscientist} evaluate each idea against the literature at generation time and keep
nothing afterwards. Our stage 9 is deliberately the weakest part of this paper --- it exists to show
that generation constrained to an offered set of catalog problems is a different operation from
free-form ideation, and that the constraint binds (104 ideas, all against offered problems).

\paragraph{Living systematic reviews.} In evidence-based medicine, a living review updates one
clinical question's conclusion as new trials appear. That is the same loop at one-question scale,
with human reviewers and a protocol. The vision here is a living review over a thousand questions
at once, which is why the human step has to be a proposal queue rather than a reviewer per problem.

\section{Conclusion}

The questions a field asks about itself --- who works on this, has it been settled, what has nobody
touched --- are queries, and today they are literature searches. We have shown that a curated
problem catalog carried unchanged under an evidence layer rebuilt from a bibliographic index turns
a useful number of them into queries, on two fields, without breaking anything the catalog-only
graph already answered. We have also shown that the most valuable query of all --- has this been
solved --- is answered by a component that gets 9 of 41 judgements right, and we would rather
publish that number than the architecture diagram alone. The design response is not a better
prompt. It is that a status change stays a proposal with its evidence attached, and that a problem
statement which names its claim is the thing that makes a refutation possible at all.

\section*{Artifact}

The pipeline, the benchmark queries, the per-question result tables and the machine-readable summary
files behind every number in this paper are at \url{https://github.com/samyama-ai/open-problems-as-data},
tag \code{v1.0-paper29}, under Apache-2.0 for the code and CC BY 4.0 for the measurements. Cite the
tag: the repository is a frozen artifact and will not carry fixes for the defects reported here.
The two catalogs and the two catalog-only graphs are separate public repositories, named with the
exact commits used in the results files; the graph engine is Apache-2.0 and runs from a container
image that pulls anonymously. No vocabulary data is redistributed --- the two vocabularies are
fetched from their publishers against recorded hashes, and the ACM classification's licence is
unstated, which the artifact says in place rather than in a footnote. The per-call run logs are not
published: they are large and hold third-party abstracts, and every number here resolves to a
committed summary file instead.

\IfFileExists{IEEEtran.bst}{\bibliographystyle{IEEEtran}}{\bibliographystyle{plain}}
\bibliography{refs}
\end{document}